\documentclass[aps,prb,twocolumn]{revtex4}
\usepackage{graphicx,amsmath,amssymb,bm}

\begin{document}

\title{Skyrmions at Vanishingly Small Dzyaloshinskii-Moriya Interaction or Zero Magnetic Field}

\author{Sandip Bera}
\affiliation{Department of Physics, Indian Institute of Technology, Kharagpur, West Bengal 721302, India}

\author{Sudhansu S. Mandal}
\affiliation{Department of Physics, Indian Institute of Technology, Kharagpur, West Bengal 721302, India}


\begin{abstract}
	By introducing biquadratic together with usual bilinear ferromagnetic nearest neighbor exchange interaction in a square lattice, we find that the energy of the spin-wave mode is minimized at a finite wavevector for a vanishingly small Dzyaloshinskii-Moriya interaction (DMI), supporting a ground state with spin-spiral structure whose pitch length is unusually short as found in some of the experiments. Apart from reproducing the magnetic structures that can be obtained in a canonical model with nearest neighbor exchange interaction only, a numerical simulation of this model with further introduction of  magnetic anisotropy and magnetic field predicts many other magnetic structures some of which are already observed in the experiments. Amongst many observed structures, nanoscale skyrmion even at vanishingly small DMI is found for the first time in a model. The model provides the nanoscale skyrmions of unit topological charge at zero magnetic field as well.We obtain phase diagrams for all the magnetic structures predicted in the model.

\end{abstract}

\maketitle

\section{Introduction}
Ever since its first realization \cite{Muhlbauer09}, the magnetic skyrmion has drawn a huge attention because of its promising applications \cite{Fert13} such as memory device, logic gate, writing and deleting information, and emergent properties \cite{Review1,Fert17} such as topological Hall effect.
The paradigm \cite{Review1,Fert17,Bogdanov} of chiral magnets describes that the nearest neighbor ferromagnetic exchange interaction, $J_1$, which favors a ferromagnetic ground state competes with the DMI energy, $D$, arising due to inversion asymmetry. As the DMI tends to make neighboring spins noncollinear, the ground state transforms into a one-dimensional spin-spiral (SS) structure with pitch length $\lambda_D = 2\pi a J_1 / D$  where $a$ is the lattice constant. The effective Zeeman energy, $H$, favors spin orientation along $+\hat{z}$ direction and thus the SS structure is converted into a skyrmion with topological quantum number $Q=1$ in the ferromagnetic background. For further increase of $H$, the size of a skyrmion becomes shorter \cite{Review1,Fert17} and eventually diminishes into the background of out-of-plane ferromagnet. The DMI of chiral magnets is quintessential for producing nanoscale SS and skyrmion structures.
The out-of-plane magnetic anisotropy,  $A$,  tends to orient all magnetic moments along perpendicular to plane and thus the SS becomes unstable for large $A$ in favor of out-of-plane ferromagnet even at $H=0$. 
 Finally, an in-plane magnetic anisotropy tends to orient the magnetic moments in a plane and by the influence of the DMI, meron lattice with $Q= 1/2$ is produced \cite{Phatak12,Yu18b,BM}
  before it becomes a planar ferromagnet. The out-of-plane tilting angle of this ferromagnet increases  till the structure becomes out-of-plane ferromagnet as $H$ is increased.

\subsection{Motivation and Model}

 In contrast to the above standard paradigm, some of the experiments report SS structures with  much shorter \cite{Meyer19,Ferriani08,Dupe14,Meckler09,Heinze11} pitch than $\lambda_D$, the skyrmions\cite{Herve18,Boulle16,Meyer19,Yu18,Huang12,Gallagher17,Zheng17,Ho19,Brandao19,Karube17,Zuo18,Desautels19} at zero magnetic field, and the skyrmions in centrosymmetric \cite{Nagao13,Yu14,Khanh2020} systems where the DMI is absent. The density functional theory (DFT) data \cite{Ferriani08,Dupe14} of spin-wave dispersion supports this short-pitched SS structure. Our aim is to introduce a theory in which all these novel magnetic structures along with the usual structures mentioned in the previous section, can be achieved in a single footing.
To this end, we consider nearest neighbor positive biquadratic exchange interaction \cite{Kittel60,Thorpe72} along with the conventional ferromagnetic exchange interaction.
On top of the above mentioned experimentally observed phases, our model predicts several other nontrivial magnetic structures that may be experimentally realized.

The presence of the biquadratic  exchange interaction in various magnetic systems has been found \cite{Hoffmann,Fedorova,Chattopadhyay,Ferriani07,Hayami17} earlier.
It generally arises due to the super-exchange mechanism \cite{Kittel60} that may also be found in the fourth order \cite{Hoffmann} of the expansion parameter $t/U$ (where $t$ is the hopping integral and $U$ is the strength of onsite Hubbard interaction) and thus usually smaller than the bilinear exchange interaction which arises in the quadratic order. However, as shown in Ref.\onlinecite{Hoffmann}, the strength of bilinear term decreases due to fourth order correction.  In the antiferromagnetic 
Fe/Ru(0001) system , 
the DFT calculation \cite{Hoffmann} finds the ratio between the strength of biquadratic and bilinear terms, $J_2/J_1 \sim 0.65$, where $J_2,\, J_1 <0$ (see Eq.\ref{Eq1} for the sign convention), indicating that the former cannot be ignored. To the best of our knowledge, no {\it ab initio} calculation has yet found the opposite signs of the exchange interactions for ferromagnetic systems, i.e.,  $J_1, \, J_2 >0$ which we on the other hand consider for our model. The microscopic calculation in Ref.\onlinecite{Hoffmann} indeed suggests that $J_2$ can be positive and $J_1$ is also positive provided $t/U \lesssim 1/2$.
We believe that the DFT calculations will reveal such a possibility specially in some of the centrosymmetric systems where skyrmions are observed. We envisage this because as these systems have very weak DMI, there must be some compensating interaction which will frustrate the system from being a ferromagnet. A model with Ruderman, Kittel, Kasuya, and Yosida (RKKY) type frustrating bilinear exchange interaction is already proposed \cite{Ferriani08,Dupe14} in literature for explaining the DFT data of non-ferromagnetic spin-wave dispersion, but it will not be appropriate for nanoscale magnetic structures as the interaction is long-ranged.  
Indeed, the importance of four-spin interaction whose two-site contribution is nothing but the biquadratic exchange interaction have been attributed to the atomic-scale skyrmions in Fe/Ir(111) system \cite{Heinze11}.


We consider a square lattice, as an illustration, of a magnetic system with Hamiltonian
${\cal H} = {\cal H}_{{\rm \small EX}} + {\cal H}_{{\rm \small D}} + {\cal H}_{{\rm \small A}} + {\cal H}_{{\rm \small Z}}$,
 where the spin-exchange Hamiltonian 
 \begin{equation}
 {\cal H}_{{\rm \small EX}} = \sum_{<ij>} \left[ -J_1 (\bm{m}_i \cdot \bm{m}_j) + J_2 (\bm{m}_i \cdot \bm{m}_j)^2 \right]
 \label{Eq1}
 \end{equation} 
 with nearest neighbor bilinear and biquadratic exchange couplings ($J_1 , J_2 >0$) and magnetization unit vector $\bm{m}_i$ at $\imath^{{\rm th}}$ site, inversion asymmetry induced Dzyaloshinskii-Moriya interaction $ {\cal H}_{{\rm \small D}} =  D\sum_{<ij>} (\hat{z} \times \hat{r}_{ij})  \cdot (\bm{m}_i \times \bm{m}_j)$ with strength $D$ for the systems with $C_{nv}$ symmetries, energy due to magnetic anisotropy ${\cal H}_{{\rm \small A}} =  A\sum_i (m_{i}^z)^2$ with positive (negative) sign of $A$ representing in-plane (out of plane) anisotropy, and Zeeman energy due to application of magnetic field $ {\cal H}_{{\rm \small Z}} = -H\sum_i m_{i}^z$ represented by energy parameter $H$ including both applied and demagnetization fields.

We find that the biquadratic exchange energy, beyond a threshold value, minimizes spin-wave dispersion energy at a nonzero momentum signaling the possibility of noncollinear magnetic structures. 
The pitch length of the SS structure is found to be {\em much} smaller than $\lambda_D$.
Employing the method of simulated annealing  by further inclusion of DMI, magnetic anisotropy and magnetic field in the model, we obtain various noncollinear magnetic structures that are experimentally observed but most of which are not yet theoretically understood for  the presence of all or the absence of one or more of these parameters: spin-spiral\cite{Meyer19,Ferriani08,Dupe14,Meckler09,Heinze11,Yu2016}; skyrmion lattice\cite{Review1,Yu18,Munzer10}; isolated skyrmions\cite{Herve18,Meyer19,Yu10,Yu2016,Romming636}; mixed phase of broken spirals, chiral bubbles, and skyrmions;\cite{Brandao19,Romming15,Yu17} meron lattice\cite{Phatak12,Yu18b,Liu}.

A naive back of the envelope calculation suggests that ${\cal H}_{{\rm \small EX}}$ is minimized when the relative orientation of two neighboring magnetic moment is $\pm \cos^{-1} (J_1/2J_2)$ for $J_1 < 2J_2$ and zero otherwise. The degeneracy in the mode of orientation (clockwise or anticlockwise) when $J_2 > J_1/2 $ is broken for an infinitesimal $D$ which favors one kind of orientation only, depending on its sign. We thus expect the spin-spiral ground state with pitch length $\lambda_{\rm \small EX} \sim 2\pi a/\cos^{-1}(J_1/2J_2)$ in the limit of vanishingly small $D$. $\lambda_{\rm \small EX}$ decreases with the increase of $J_2/J_1$ and it's  about $11$ lattice sites for $J_2 = 0.6J_1$. It can further decrease with the increase of $D$. We note that even for vanishingly small DMI relevant for centrosymmetric systems, SS structure with pitch-length of a few lattice sites is possible. For the same reason, we expect, contrary to the paradigm, the formation of small size skyrmions even in centrosymmetric systems but with $J_2 >J_1/2$ when magnetic field is applied.
Does this model also favor stabilizing skyrmions in the absence of magnetic field?  
 $J_1$ supports $J_2$ for orienting neighboring spins along same direction, but it cannot break symmetry to orient all spins along a particular direction because  $J_2$ will destabilize.  However, if large easy-axis magnetic anisotropy spontaneously creates up (down) spin background then the spiral effect generated by the exchange interactions can produce skyrmions with topological number $Q=1$ whose center will have down (up) spin moment. 
 We next describe our detailed results obtained using analytical calculation and numerical simulation, that will be in agreement with the above intuitive description.

 	
 	\begin{figure}[h]
 		\begin{center}
 			\includegraphics[scale=0.75]{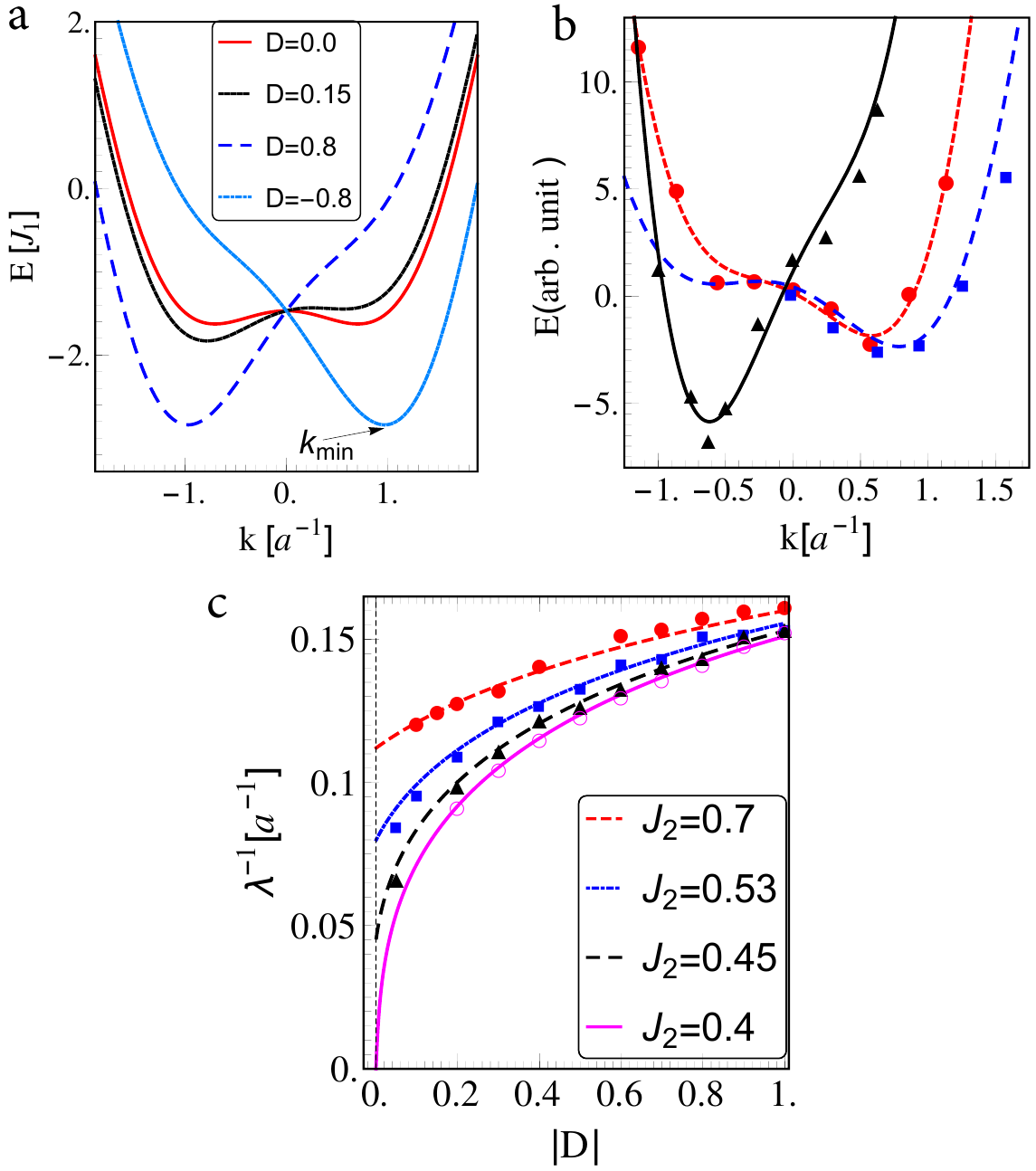}
 			\caption{ 
 				(a) 
 				The dependence of the spin-wave energy (2) on wave number for $J_2=0.7$. 
 				$k_{{\rm min}}$, the wave number corresponding to the minimum energy shifts to the higher magnitude with the increase of $D$ and its sign is opposite to the sign of $D$. (b) Dispersion obtained in {\it ab-initio} calculations for thin films Pd/Fe/Ir\cite{Dupe14} (circles), 2Pd/Fe/Ir\cite{Dupe14} (squares), and in Mn/W\cite{Ferriani08}  (triangles) are fitted (red, blue, and black lines respectively) with dispersion relation in Eq.~(2). Energies are appropriately scaled (multiplied by a factor) for plotting together. The extracted parameters are shown in Table-1.  (c) Inverse of the pitch-length $\lambda^{-1}\equiv k_{{\rm min}}/2\pi$ {\it vs.} $\vert D\vert$. 
 				The symbols closest to these lines represent the corresponding  values obtained from our simulation in a $128 \times 128$ square lattice.}
 		\end{center}	
 	\end{figure}


  \section{Spin-wave Dispersion}
  
   We find (see Appendix A for details) the spin-wave  dispersion from the Hamiltonian ${\cal H}_{\rm EX} + {\cal H}_{\rm D}$ for $k_y=\pm k_x$ (along high-symmetry directions)  as
   \begin{equation}
    E_{\rm k} = -4 \left( 1+ \frac{16}{\pi^2} J_2\right) \cos(k_x) + 4J_2 \cos(2k_x)  
     + 4D \sin (k_x)  \, .
     \label{dispersion}
   \end{equation}
   Henceforth, we assume all length scales are in the unit of  $a$ and energy scale in the unit of $J_1$.
   Figure 1{\bf a} shows $E_{\rm k}$ for  $D=0,\,0.15,\,0.8,\,-0.8$ and $J_2=0.7$.  We find two degenerate minima at two nonzero $k_x$ (equal in magnitude but opposite in sign) for $D=0$. For any arbitrary nonzero magnitude of $D$, this degeneracy is broken and one global minimum occurs at $k_x = -k_{\rm min} \,{\rm sgn}(D)$, and $k_{\rm min}$ increases with $\vert D\vert$. Energy dispersion obtained in {\em ab initio} studies \cite{Ferriani08,Dupe14} fit (Fig.1{\bf b}) quite well with the analytical form in Eq.(\ref{dispersion}). With the input value of $J_1$ provided in these {\em ab initio} studies , we extract $J_2$, $D$ and $\lambda = 2\pi /k_{\rm min}$ which are tabulated in Table~1. The pitch length $\lambda$ of SS excellently agrees with experiments \cite{Ferriani08,Dupe14} on  Mn/W and Pd/Fe/Ir systems. Figure 1{\bf c} shows the variation of $\lambda^{-1}$ with $\vert D \vert$. We note that while $\lambda^{-1}$ is zero at $D=0$ for $J_2 = 0.4$, it is nonzero for $J_2 = 0.45$. To be precise, SS structure is possible even at $D=0$ for $ 0.42 \lesssim J_2$; the lower bound of $J_2$ for SS structure is somewhat less than the naive calculation discussed above.
   As $J_2$ is increased, the dependence of $\lambda$ on $D$ decreases and it becomes shorter. Therefore, the energy scale $J_2$ provides SS structures with shorter pitch length than the same for the presence of $D$ only. For example, if $J_1=1$ and $D=0.05$, the standard paradigm supports spin-spiral pitch length of about 125 lattice sites, while our model with $J_2 = 0.5$ keeping the same values of $J_1$ and $D$ supports spin-spiral of pitch length 12 lattice sites.
   However, if $J_2 < 0.42$, like the canonical model, large DMI is essential to
   produce spin spiral of similar pitch-length. 
   Therefore, the present model with $J_2>0.42$ has clear advantage (which we show below) over the canonical model for explaining nanoscale magnetic structures for the physical systems with small DMI, for example, thin films made with centrosymmetric crystals. 
     
 \begin{table}
 	\caption{Model Parameters extracted from  {\em ab initio} Studies.  The values of $J_1$ and $a$ are taken from {\em ab initio} studies; $D/J_1$, $J_2/J_1$ and $\lambda= 2\pi/k_{\rm min}$ are extracted from fitting of spin-wave dispersion obtained in these studies by the formula in Eq.(\ref{dispersion}). The corresponding experimental values of $\lambda$ are also tabulated.  }
 	\begin{tabular}{|c|c|c|c|c|c|c|}\hline\hline
 		System &  $J_1$ & $a$ & $D/J_1$ & $J_2 /J_1$ & $\lambda $ [nm] & $\lambda$ [nm]\\
 		&[meV] & [nm]& &  &{\rm Theory} & {\rm Expt} \\
 		\hline
 		Mn/W\cite{Ferriani08} &    19.7&0.22 & 0.15 & 0.48 & 2.23 & 2.2 \\
 		\hline
 		Pd/Fe/Ir\cite{Dupe14} (fcc) &  14.7& 0.27 & -0.09 & 0.51 & 2.91 & 3.0  \\ \hline
 		2Pd/Fe/Ir \cite{Dupe14} &  9.0 & 0.29 & -0.15 & 0.71 & 2.29 & 2.3\\ \hline
 	\end{tabular}
 \end{table}
 
 Alternatively, the long ranged (more than $6^{{\rm th}}$ nearest neighbor) exchange couplings \cite{Ferriani08,Dupe14} which may arise due to long-ranged RKKY type interaction seem to reproduce this short pitch of spin-spiral. The model with bilinear and biquadratic nearest neighbor exchange interaction may be as relevant as the model with long-ranged bilinear exchange interaction to the systems that have been studied by these authors. While both the models describe spin-spiral for vanishingly small DMI, it distinguishes the models as the former (latter) provides equal (unequal) change in spin orientation between two neighbors that may easily be understood through their respective continuum versions. However, such a model with  long-ranged and multi-parameter exchange couplings is possibly not relevant when the skyrmions produced due to the superposition of spin spirals along high symmetric directions is much shorter in size compared to this  range of the interaction. 
 Further, exchange interaction up to second nearest neighbor (a short-range version of RKKY type interaction) seems to suggest skyrmions with $Q>1$, instead of $Q=1$ \cite{Leonov15}.
 
 \begin{figure}[h]
 	\includegraphics[scale=0.82]{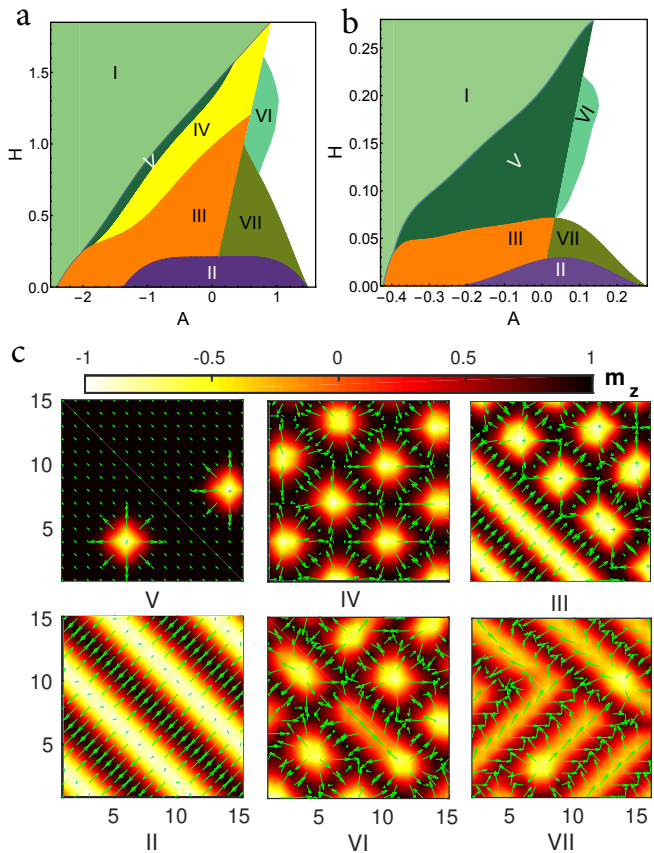}	
 	\label{Phase1}
 	\caption{ 
 		 Phase diagrams for a high DMI, $D=0.8$ (a) and for a relatively  low value of DMI, $D=0.15$ (b) in $H$--$A$ plane obtained by simulation in a $32 \times 32$ square lattice (a quarter of it is shown for clarity) with periodic boundary condition and $J_2=0.7$. 
 		 The magnetic phases shown in (a) and (b) are  {I}: Out-of-plane ferromagnet, {II}: cycloidal spin-spiral, {III}: mixed phase of broken spin-spiral, skyrmions, and chiral bubbles, {IV}: skyrmion lattice, {V}: isolated skyrmions, {VI}: The mixed phase of skyrmions and elongated magnetic vortex with one-dimensional elongation, { VII}: mixed phase of broken spin spiral and elongated magnetic vortex with one-dimensional elongation. The phase corresponding to white region is unclear and some of the corresponding magnetic structures are shown in supplemental as illustration. (c) Magnetic structures obtained for noncollinear phases ({ II -- VII}). The arrows indicate the inplane component of magnetization while the out-of-plane component of  has been color coded.}		
 \end{figure}
 
\section{Numerical Simulation and Results}

 We next perform numerical simulation (see Appendix B for its method) of ${\cal H}$ for determining ground state magnetic structures in the parameter space of $A$, $D$ and $H$ for $J_2= 0.7$ which supports SS even for small $D$ as discussed above. Various magnetic phases obtained in the simulation are characterized by the estimation of following parameters: local magnetization $m_i^z$, average magnetization $M = (1/N)\sum_i m_i^z$, local chirality $\rho_i = (1/4\pi) \sum_{\delta=\pm 1} \bm{m}_i \bm{\cdot} \left( \bm{m}_{i+\delta\hat{x}} \times \bm{m}_{i+\delta\hat{y}} \right)$, and total chirality ${\cal R} = (1/N)\sum_i \rho_i$; local spin asymmetric parameter $\delta\theta_i = \vert \theta_{i+\hat{y}}-\theta_{i-\hat{y}}\vert -\vert \theta_{i+\hat{x}}-\theta_{i-\hat{x}}\vert$, where $N$ is the number of lattice sites. We find various magnetic phases (I--XI as marked in Figs.~2 and 3) characterized by different  ranges of these parameters shown in Table-II.  
 
 \subsection{Structures for Large DMI}
  
 Figure 2{\bf a} shows phase diagrams in $H$--$A$ plane for a larger DMI, $D=0.8$. As expected from the paradigm of chiral magnets, we obtain polarized ferromagnet (phase-I), spin-spiral (phase-II), isolated skyrmions (phase-V)\cite{footnote}, and skyrmion lattice (phase-IV) phases. The pitch of the skyrmion lattice strongly depends on $J_2$ (about 6 (13) atomic lattice sites for $J_2=0.7 \, (0.3))$.
 While this pitch is almost independent of $D$ when $J_2$ is large, it depends on $D$ when $J_2$ is very small, in agreement with the simulations \cite{Siemens2016,Fattouhi2011} for the model with bilinear exchange term only. 
 The polarized ferromagnet is obtained for $A\lesssim -2.4$ at $H=0$ and thereafter for lowering out-of-plane anisotropy (higher $A$ including its sign) with the increase of $H$. The SS phase is obtained for $\vert A\vert \lesssim 1.4$. 
 The skyrmion phases are obtained when $0.22 \lesssim H$.

  In addition to these phases, we obtain a mixed phase of broken spirals, skyrmions, and chiral bubbles  (phase-III in Fig.~2) mainly for  a range of out-of-plane magnetic anisotropy and small values of in-plane anisotropy. This phase occurs for a medium range in $H$ by separating  SS and skyrmion lattice, and it is also extended up to $H=0$ for $ -2.4 \lesssim A \lesssim -1.2$. 
 Two other new structures have been found for in-plane anisotropy and medium range of $H$: elongated magnetic vortices (having fractional skyrmion number) with their extension along high-symmetry directions (phase-VII in Fig.~2) of the underlying lattice at relatively lower $H$ but beyond the SS phase; a mixed phase (phase-VI in Fig.~2) of skyrmions and elongated magnetic vortices (having fractional skyrmion number) at relatively higher $H$. All the non-collinear structures obtained in the simulations are illustrated in Fig.~2{\bf c}. 
 
 \begin{table}
 	\caption{Characterization of magnetic phases in terms of various parameters calculated from the ground states.
  While all the parameters sometimes may be same for the phases X and XI, they can be distinguished through the Fourier decomposition as the former is an ordered structure and the latter is disordered.  }
 \begin{tabular}{|c|c|c|c|c|c|}\hline\hline
 	Phase & $m_i^z $ & $M$ & $\rho_i$ & ${\cal R}$ & $\delta\theta_i$ \\  \hline
 	I & $\lesssim 1$ & $>0.98$ & $\sim 0.0$ & $<0.001$ & $0$ \\ \hline
 	II & (-1)--(+1) & $<0.001$ & $\sim0.0$ & $<0.001$ & $\neq 0$ \\ \hline
 	III & (-1)--(+1) & 0.1--0.6 & 0.0--1.0 & 0.25--0.45 & $\neq 0$ \\ \hline
 	IV & at the core  &  &at the core   &   &  \\
 	 & (-0.98)--(-1.0) &0.45--0.8 &0.95--1.0 &0.5--0.9 & 0\\  \hline
 	 V & at the core  &  &at the core  &   &  \\
 	 & (-0.98)--(-1.0) &0.8--0.98 &0.95--1.0 &0.04--0.5 & 0\\  \hline
 	 VI & at the core&0.3--0.45& 0.95--1.0&0.5--0.7 & $\neq 0$\\ 
 	  & (-0.98)--(-1.0) & & & & \\ \hline
 	 VII &(-1)--(+1) &0.04--0.16 & $<0.95$&$<0.4$ & $\neq 0$\\ \hline
 	 VIII & at the core & $\sim 0.001$ & $\sim 0.5$ & $\sim 0.0$ & 0 \\ 
 	 &$\vert m_i^z\vert>0.98$  & & & & \\ \hline
 	 IX & $\sim 0.5$ & $\simeq 0.5$ & $\simeq 0.0$  & $<0.001$ & 0 \\ \hline 
 	 X& (-0.95)--(0.95) & Any Value & $<0.001$ & $<0.001$ &  $\neq 0$ \\ \hline
 	  XI& (-0.95)--(0.95) & $<0.001$ & $<0.001$ & $<0.001$ & $\neq 0$  \\ \hline 
 	\hline 
 \end{tabular}	
 \end{table}

\subsection{Structures for Vanishingly Small DMI}
 
 As we decrease the value of $D$, say for $D=0.15$, all the above phases excepting the skyrmion lattice phase are obtained (Fig.~2{\bf b}). They, however, occur, for different ranges of $A$ and $H$. One interesting finding is that isolated skyrmions with radius of few lattice sites are also possible at such a low value of $D$. For further lowering the value of $D$, we have  extended our simulation for the system size up to $64 \times 64$ and have found that isolated skyrmions are possible for as low as $D=0.05$. We thus conclude that our model ${\cal H}_{{\rm EX}}$ supports isolated skyrmions for vanishingly small DMI, in contrast to the paradigm of chiral magnets, at the thermodynamic limit. 
 This is corroborated with the experiments \cite{Nagao13,Yu14,Herve18,Khanh2020} that have reported skyrmions in the centrosymmetric systems.
 Yu et al \cite{Yu2016} have interestingly found our proposed phases II, III, V, I (see Fig.~2b) which are possible for negative $A$ and small $D$ by varying magnetic field.
 The essential breaking of degeneracy in the minima of spin wave dispersion is presumably done
 by weak DMI produced in thin films \cite{Bogdanov2001,Grigoriev2008} made with centrosymmetric bulk systems. These systems may also have dipolar interactions \cite{Fert17,Ezawa10}.
 However, unlike our findings, neither dipolar interaction nor weak DMI can stabilize magnetic structures with nanoscale (a few nanometer) size.

 \begin{figure}[h]
 	\includegraphics[scale=0.92]{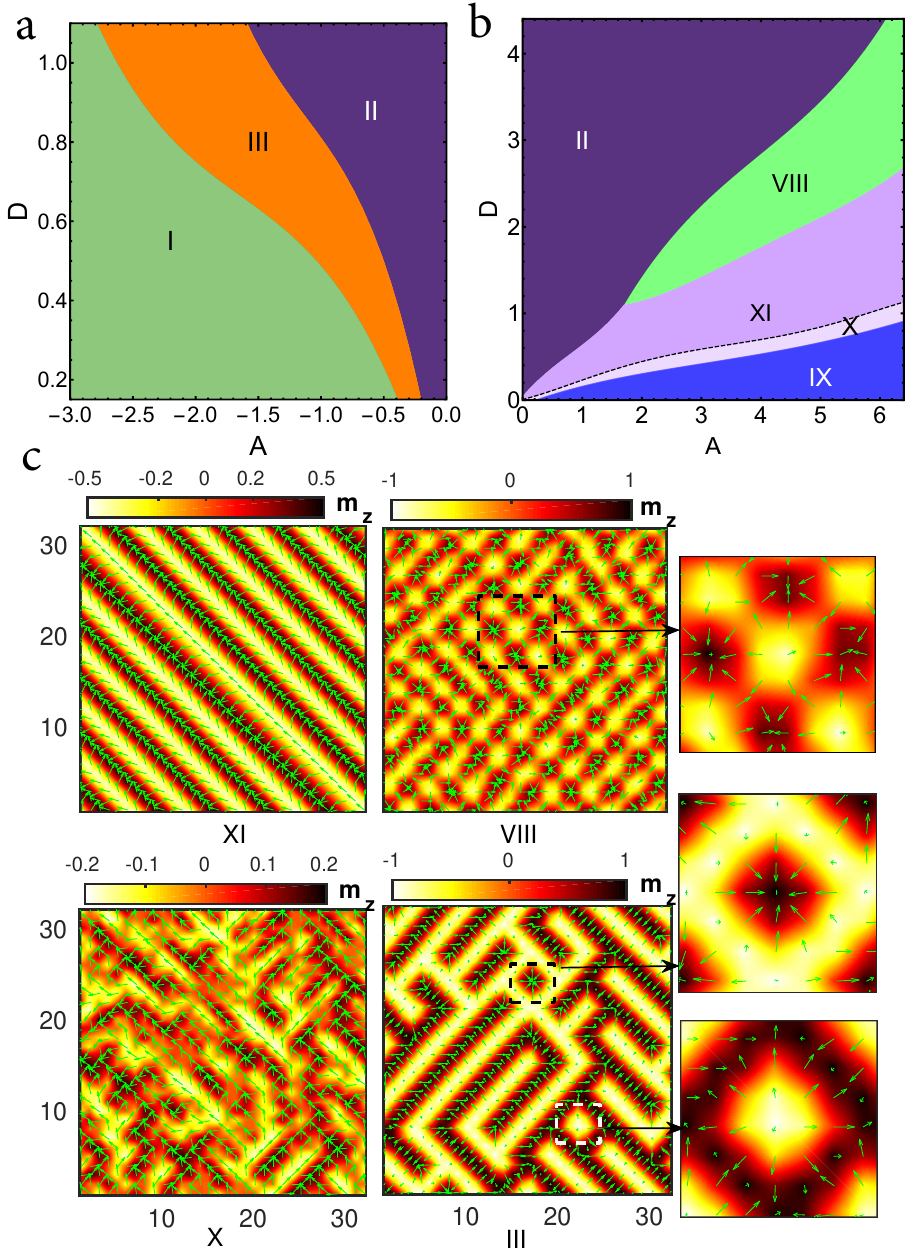}	
 	\label{Phase2}
 	\caption{ 
 		Phase diagrams in $D$--$A$ plane for out-of-plane magnetic anisotropy (a) and for  easy-plane anisotropy (b) obtained by simulation in a $32 \times 32$ square lattice with periodic boundary condition and $J_2=0$ at zero magnetic field.
 	 The new pases in comparison to Fig.~2 are { VIII}: meron lattice; { IX}: planar ferromagnet; { X}: random spin islands of positive and negative out-of-plane magnetization; { XI}: spin-spirals without complete $2\pi$ spin rotation.   (c) Magnetic structures obtained for unconventional phases (XI, VIII, X, and III).
 	 Zoomed structures from top: Meron lattice, skyrmion with spin-up center, and skyrmion with spin-down center.   }		
 	
 \end{figure}

 \subsection{Structures at Zero Field}   
 
 We next analyze magnetic structures at $H=0$. Figs.~3{\bf a} and 3{\bf b} respectively show the corresponding phases for $A<0$ and $A>0$ in $D$--$A$ plane. Interestingly, apart from expected polarized ferromagnet and SS structures, a mixed phase of broken spin-spirals, chiral bubbles and skyrmions (phase-III) are found for out-of-plane magnetic anisotropy. These skyrmions are likely to be corroborated with the skyrmions observed at zero magnetic field at various experiments.\cite{Herve18,Boulle16,Meyer19,Yu18,Huang12,Gallagher17,Zheng17,Ho19,Brandao19,Karube17} Meyer et al \cite{Meyer19} have recently reported that skyrmions at zero magnetic field are observed only for the systems with energy dispersion having quartic wavenumber dependence rather than quadratic only. However, the dispersion relation with positive coefficients for  quadratic  term will not make any qualitative difference. The sign of quadratic and quartic terms ought to be negative and positive respectively. Our dispersion relation (\ref{dispersion}) for $D=0$
 at small momenta provides exactly this expectation. The dispersion relation at small momenta reads
 \begin{eqnarray}
 E_ {\rm k} &=& 4\left[-1+J_2\left(1-\frac{16}{\pi^2}\right) \right]+ \left[ 2-8J_2\left(1-\frac{4}{\pi^2}\right)\right]k_x^2 \nonumber \\
 && + \frac{1}{6}\left[ -1+ 16J_2\left( 1-\frac{1}{\pi^2} \right)\right]k_x^4
 \end{eqnarray}
 which suggests negative coefficient for quadratic term when $J_2 >0.42$, which is consistent with our detailed analysis about the lower bound of $J_2$ in section II. 
  

  We find that the skyrmions with down as well as up spin at their centers (see zoomed structures of phase-III in Fig.~3{\bf c}) may simultaneously form in the respective background of up and down spins because both these backgrounds at zero magnetic field are degenerate. In the case of positive anisotropy, apart from the well-known phases of SS and planar ferromagnet, we find a regime of $D$ and $A$ that supports a lattice of merons--a meron with up spin at its center is surrounded by merons with down spin at their centers and vice versa (phase-VIII in Fig.~3{\bf c}). This lattice has been recently reported in  experiments \cite{Phatak12,Yu18b,Liu}, a micromagnetic simulation \cite{Yu18b} and also as analytic  solution \cite{BM} of Euler equation of a single meron followed by physical arguments. For the parameter regime in between planar ferromagnetic and meron lattice phases, we find a SS structure like phase but with the limited range ( phase-XI in Fig.~3{\bf c}) of out-of-plane magnetization, $\vert m_z \vert < 1$. The value of $\vert m_z \vert$ increases as we approach closer to meron lattice boundary. However, we find a narrow regime of parameter space towards the planar ferromagnet phase boundary where the structure appears to be the disordered spin island (phase-X in Fig.~3{\bf c}) phases with $\vert m_z\vert <1$.

\section{Analysis of the Structures}
   
 The pitch length of SS structure obtained in this simulation method for various values of $J_2$ agrees very well with the analytical estimate for an infinite system (Fig. 1{\bf c}). For increasing accuracy in estimating $\lambda$ from the simulation for lower $\vert D\vert$, we have considered $128 \times 128$ lattice. The minimum value of $\vert D\vert$ for which SS structure is obtained for a given lattice size is at the threshold value of $J_2 \approx 0.42$ which agrees with its analytical estimation.

 The spin configuration of some of the magnetic phases may be represented in terms of the truncated Fourier decomposition\cite{Leonov15,Gastiasoro,Seabra}: 
 \begin{equation}
 \bm{S}_i = \bm{B}_0 + \frac{1}{2}\sum_{\alpha =1}^{2} \left( \bm{B}_\alpha e^{i \bm{q}_\alpha \bm{\cdot} \bm{r}_i} + \bm{B}^\ast_\alpha e^{-i \bm{q}_\alpha \bm{\cdot} \bm{r}_i} \right)
 \end{equation}
 where $\bm{q}_{1,2} = (q/\sqrt{2})(1,\,\pm 1)$ are two orthogonal wave vectors representing two high symmetric directions in a square lattice. The uniform ferromagnetic state is represented by $B_0^z =1$ and $B_0^x=B_0^y =\bm{B}_\alpha = 0$. We find that the SS structure corresponds to $\bm{B}_0 = \bm{B}_2=0$ and $\bm{B}_1 = (-i/\sqrt{2},\, -i/\sqrt{2},\, 1)$. The direction of propagation of the spin-spiral is along $\bm{q}_1$ which switches over to $\bm{q}_2$ when the sign of DMI changes. Superposition of modulations along both $\bm{q}_\alpha$ directions with equal amplitude creates the skyrmion structures with circular symmetry. In this case, $\bm{B}_0 = 0$, $\bm{B}_1 = B e^{i\phi_1}(-i \sin \chi, i \cos\chi,1)$ and $\bm{B}_2 = B e^{i\phi_2}(-i \cos \chi, -i \sin\chi,1)$ where $B=-1/2$ satisfying $S_i^z = -1$, $S_i^x=S_i^y=0$ at the center of the skyrmions. The magnetic vortices are also formed due to the superposition of both $\bm{q}_1$ and $\bm{q}_2$ but with unequal amplitudes.

\section{Discussion and Conclusion}
 
By introducing biquadratic nearest neighbor exchange interaction, we comprehensively show numerous exotic magnetic structures such as spin-spiral with unusually short pitch-length, skyrmions at zero magnetic field and/or vanishingly small Dzyaloshinskii-Moriya interaction, and meron lattice at zero magnetic field. We thus have made a compelling case of explaining a large number of recent experiments observing these unusual magnetic structures in a single model. Our phase diagrams will encourage further investigations of several other magnetic structures including these over their respective ranges of parameter space. 
Our theory should also be relevant for recent observation \cite{Morin16} of spin-spiral in YBaCuFeO$_5$  with extremely small Dzyaloshinskii-Moriya interaction \cite{Dibyendu18}. We predict that these systems should also host skyrmions and other magnetic structures shown in the phase diagrams.
 
Our study is based on the model consisting of nearest neighbor ferromagnetic bilinear and a positive biquadratic exchange interactions. {\em Ab initio} studies are indeed necessary for investigating such interactions in some of the  systems, especially the centrosymmetric systems, where skyrmions are observed.


    
\appendix	

\section{Energy Dispersion} 

In this appendix, we show detailed calculation of the spin-wave dispersion relation for the exchange Hamiltonian: 
\begin{equation}
	{\cal H}_{{\rm \small EX}} = \sum_{<ij>} \left[ -J_1 (\bm{m}_i \cdot \bm{m}_j) + J_2 (\bm{m}_i \cdot \bm{m}_j)^2 \right].
	\label{H-EX}
\end{equation}
By employing the method discussed in Ref.\onlinecite{Kittel63}, we write $m_j$'s in terms of Bosonic creation and annihilation operators $(a_j^\dagger , a_j)$ with the commutation relation $[a_j,a_l^\dagger] = \delta_{jl}$ as $m_j^+ = m_{j}^x + im_j^y = (2S - a_j^\dagger a_j)^{1/2}a_j$, $m_j^- = m_{j}^x - im_j^y = a_j^\dagger (2S - a_j^\dagger a_j)^{1/2}$, and $m_j^z = S-a_j^\dagger a_j$ where the magnitude of spin $S$ is assumed to be large for perturbative expansion and finally we take the limit $S \to 1$. As we substitute $\bm{m}_i$ in Eq.(\ref{H-EX}) by the above Bosonic operators, $\bm{m}_i \bm{\cdot} \bm{m}_j$ should be replaced by $\bm{m}_i \bm{\cdot} \bm{m}_j -S^2$.

Expressing $a_j = \frac{1}{\sqrt{N}} \sum_{\bm{k}} e^{-i\bm{k}\cdot \bm{r}_j}b_{\bm{k}}$ and $a_j^\dagger = \frac{1}{\sqrt{N}} \sum_{\bm{k}} e^{i\bm{k}\cdot \bm{r}_j}b_{\bm{k}}^\dagger $ in Fourier representation with $[b_{\bm{k}},\, b_{\bm{k}'}^\dagger]= \delta_{\bm{k}\bm{k}'}$, we find
${\cal H}_{{\rm \small EX}} = H_0 + H_1 $ (ignoring terms without Bosonic operators) where 
\begin{eqnarray}
	H_0 &=& -2J_1 S \sum_{\bm{k}} (\cos (k_xa) + \cos(k_ya) -2) b_{\bm{k}}^\dagger b_{\bm{k}} \nonumber \\
	&=& -2J_1S\sum_{\bm{k}} \zeta (k) b_{\bm{k}}^\dagger b_{\bm{k}}
\end{eqnarray}
with $\zeta (k) = \cos (k_xa) + \cos(k_ya) -2$
and  
\begin{equation}
	H_1 = \frac{S^2}{N} \sum_{\bm{k},\bm{k}_1,\bm{k}'} {\cal J}(\bm{k},\bm{k}_1,\bm{k}') b_{\bm{k}}^\dagger b_{\bm{k}_1} b_{\bm{k}'}^\dagger b_{\bm{k}-\bm{k}_1+\bm{k}'}.
\end{equation}

\begin{figure}[h]
	\includegraphics[scale=0.2]{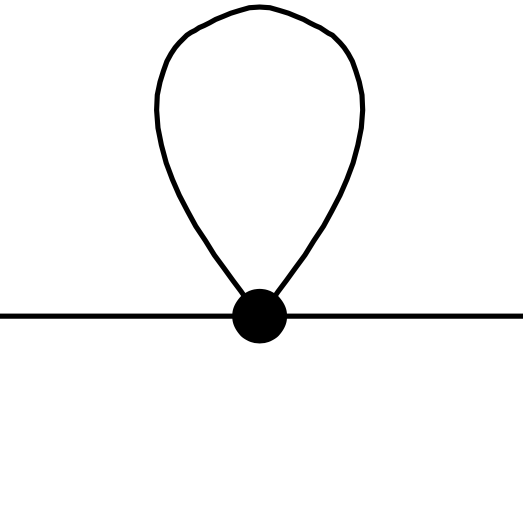}
	\vspace{-1.0cm}
	\caption{Interaction vertex of four Bosons are represented by the filled circle  with coupling strength ${\cal J}(\bm{k},\bm{k}')$. The loop indicates ensemble average  $\langle b_{\bm{k}'}^\dagger b_{\bm{k}'}\rangle$. }
\end{figure}

For the purpose of determining dispersion relation, we set $\bm{k}=\bm{k}_1$ and thus 
\begin{equation}
	\tilde{H}_1 = S^2 \sum_{\bm{k},\bm{k}'} {\cal J}(\bm{k},\bm{k}') b_{\bm{k}}^\dagger b_{\bm{k}} b_{\bm{k}'}^\dagger b_{\bm{k}'}
\end{equation}
with 
\begin{eqnarray}
	{\cal J}(\bm{k},\bm{k}') 	&=& 4J_2 \left[ \cos (k_xa)\cos(k'_xa) 
	+ \cos (k_ya)\cos(k'_ya)\right]  \nonumber \\
	&&  -(4J_2-J_1/2)\left[ \cos(k_xa) +\cos(k_ya)  \right.  \nonumber \\ &&\left. +\cos(k'_xa)+\cos(k'_ya)  \right] +(8J_2-2J_1) 
\end{eqnarray}
We thus obtain an effective Hamiltonian (quadratic in Bosonic field) as 
\begin{equation}
	H_{\rm EX}^{{\rm eff}}  = \sum_{\bm{k}} \tilde{{\cal J}}(k) b_{\bm{k}}^\dagger b_{\bm{k}}
\end{equation}
where 
\begin{equation}
	\tilde{{\cal J}}(k) = -2J_1 \zeta (k) +  \sum_{\bm{k}'} {\cal J}(\bm{k},\bm{k}')\langle  b_{\bm{k}'}^\dagger b_{\bm{k}'} \rangle
\end{equation}
by putting $S=1$. 
The ensemble average $\langle b_{\bm{k}'}^\dagger b_{\bm{k}'} \rangle$ (see Fig.~4 schematically) may be obtained using the expression of $H_0$. To that end, $\zeta (k)$ has 4-fold symmetry in the Brillouin zone, with one of such independent regime is $ 0 \leq (k_x,\, k_y)\leq \pi$ in which $\zeta (k) > \,(<) \,0$, which we denote as $\zeta^+\, (\zeta^-) $, for $k_x +k_y  \leq \, (>)\, \pi$. Therefore, 
\begin{eqnarray}
	\tilde{{\cal J}}(k) &=& -2J_1 \zeta (k) + \frac{1}{\pi^2}\left[ \int_0^{\pi} dk'_x\int_0^{\pi-k'_x} dk'_y \sum_n \frac{{\cal J}(\bm{k},\bm{k}')}{i\Omega_n - \zeta^-} \right. \nonumber \\
	& & + \left. \int_0^{\pi} dk'_x\int_{\pi-k'_x}^\pi dk'_y \sum_n \frac{{\cal J}(\bm{k},\bm{k}')}{i\Omega_n - \zeta^+} \right]
\end{eqnarray}
where $\Omega_n $ is Bosonic Matsubara frequency. We hence find energy dispersion 
\begin{eqnarray} \tilde{{\cal J}}(\bm{k}) &=& -2\left( J_1 + (16/\pi^2)J_2 \right) (\cos (k_x a) + \cos(k_y a)) \nonumber \\
	&& + 2J_2 (\cos (2k_x a) + \cos(2k_y a)) 
\end{eqnarray}
at zero temperature, by dropping the constant terms.
We note that $\tilde{{\cal J}}(\bm{k})$ has four-fold symmetry in the Brillouin zone as the energy is same at $(\pm k_x, \, \pm k_y)$. We thus find effective exchange interaction: ${\cal H}_{{\rm EX}}^{{\rm eff}} = \sum_{\bm{k}}  \tilde{{\cal J}}(\bm{k}) b_{\bm{k}}^\dagger b_{\bm{k}} \equiv \sum_{\bm{k}} \tilde{{\cal J}}(\bm{k}) \bm{m}_{\bm{k}} \cdot \bm{m}_{-\bm{k}}$,
and DM interaction energy
\begin{eqnarray}
	{\cal H}_{{\rm DM}} &=& i\sum_{\bm{k}}  \left[ {\cal D}_x (\bm{k}) 
	(m^z_{\bm{k}}m^x_{-\bm{k}} - m^x_{\bm{k}}m^z_{-\bm{k}}) \right. \nonumber \\
	&&\left. - {\cal D}_y (\bm{k}) (m^z_{\bm{k}}m^y_{-\bm{k}} - m^y_{\bm{k}}m^z_{-\bm{k}} )\right] \label{eq2}
\end{eqnarray}
where $ {\cal D}_x (\bm{k}) = 2D \sin (k_xa)$, $ {\cal D}_y (\bm{k}) = 2D \sin (k_ya)$, and $m_{\bm{k}}$ is Fourier transform of $m_i$.

Without loss of generality, we assume $\bm{k} = k_x \hat{e}_x$ and considering the variation of magnetization in the $z-x$ plane, we find the effective Hamiltonian as
\begin{equation}
	\tilde{{\cal H}} = \sum_{k_x} \left( \begin{array}{cc}  m^z_{\bm{k}},  & m^x_{\bm{k}} \end{array} \right) \left[ \begin{array}{cc}
		\tilde{{\cal J}}(k_x) & i {\cal D}_x (k_x) \\
		-i {\cal D}_x (k_x) & \tilde{{\cal J}}(k_x) \end{array} \right] \left( \begin{array}{c} m^z_{-\bm{k}} \\ m^x_{-\bm{k}} \end{array} \right)
\end{equation}
leading to the energy dispersion 
\begin{equation}
	E^\pm (k_x) = \tilde{{\cal J}} (k_x) \pm {\cal D}_x (k_x) \, .
\end{equation}	
Clearly, $E^+ (D,k_x)= E^- (-D,k_x) = E^+ (-D,-k_x)$. It is thus sufficient that we consider $E^+ (D,k_x) \equiv E(k_x)$ only.\\

\section{Simulation Method}

We obtain magnetic phase diagrams by performing simulated annealing from a large temperature for the spin model described by ${\cal H}$ in a square lattice  $32 \times 32$ size with periodic boundary condition. Some of the key results have been reconfirmed for $128 \times 128$ size also. The simulation is carried out by adopting standard Metropolis algorithm for updating local spins up to $4\times 10^6$ steps for each temperature. We gradually reduce the temperature in each step of the simulation \cite{Hayami18} following the relation $T_{n+1}=\alpha T_n$ where $T_n$ is the temperature in the $n^{{\rm th}}$ step. We set the initial temperature $T_0 = 11$ and $\alpha = 0.99$ to reach final temperature $T\sim 5\times 10^{-4}$ in $10^3$ steps. \\

{\bf Supplementary Information}\\

See supplementary material, where  we provide 
movies for the evolution of different phases in the phase diagrams (Fig.2 and Fig.~3), and show some of the unusual magnetic structures which may have close resemblance with the previously predicted asymmetric skyrmions \cite{BM}.\\





\end{document}